\documentclass[aps,twocolumn,showpacs]{revtex4}
\usepackage{graphicx}

\DeclareGraphicsExtensions{.png}

\begin{document}

\title[Short Title]{Self-organizing approach for
  automated gene identification in whole genomes}

\author{Alexander N. Gorban$^{1,2}$}
\author{Andrey Yu. Zinovyev$^{1,2}$}
\email{gorban@icm.krasn.ru}
\author{Tatyana G. Popova$^{1}$}
\affiliation{$^1$  Institute of Computational Modeling RAS, 660036
Krasnoyarsk, Russia}
\affiliation{$^2$ Institut des Hautes Etudes
Scientifiques, France}
\date{\today}

\begin{abstract}
An approach based on using the idea of distinguished coding phase
in explicit form for identification of protein-coding regions in
whole genome has been proposed. For several genomes an optimal
window length for averaging GC-content function and calculating
codon frequencies has been found. Self-training procedure based on
clustering in multidimensional space of triplet frequencies is
proposed.
\end{abstract}


\maketitle

Most of the computational approaches for identification of coding
regions in DNA have following limitations \cite{Claverie}: they
need a training set of already known examples of coding and
non-coding regions, they work with a comparably short subsequence
of DNA rather than whole sequence and they are able to recognize
mainly protein-coding regions.

Recently some approaches appeared which promise to be free of
these limitations. In the works by Yeramian E.
\cite{Yeramian00,Yeramian00-1} DNA sequence is considered as a
linear chain of strong (GC-bond) and weak (AT-bond) hydrogen
bonds. Applying a kind of Ising model to the calculation of
partition function one can obtain a thermal DNA stability map (a
plot of probability of every DNA basepair to be disrupted). With
appropriate temperature chosen, the map in some cases shows
believable correlation with the arrangement of coding regions in
DNA. This fact was exploited with some success to identify coding
regions in Plasmodium falsiparum in some non-standard for
gene-finders situations.

\begin{figure}
\includegraphics[width=70mm, height=50mm]{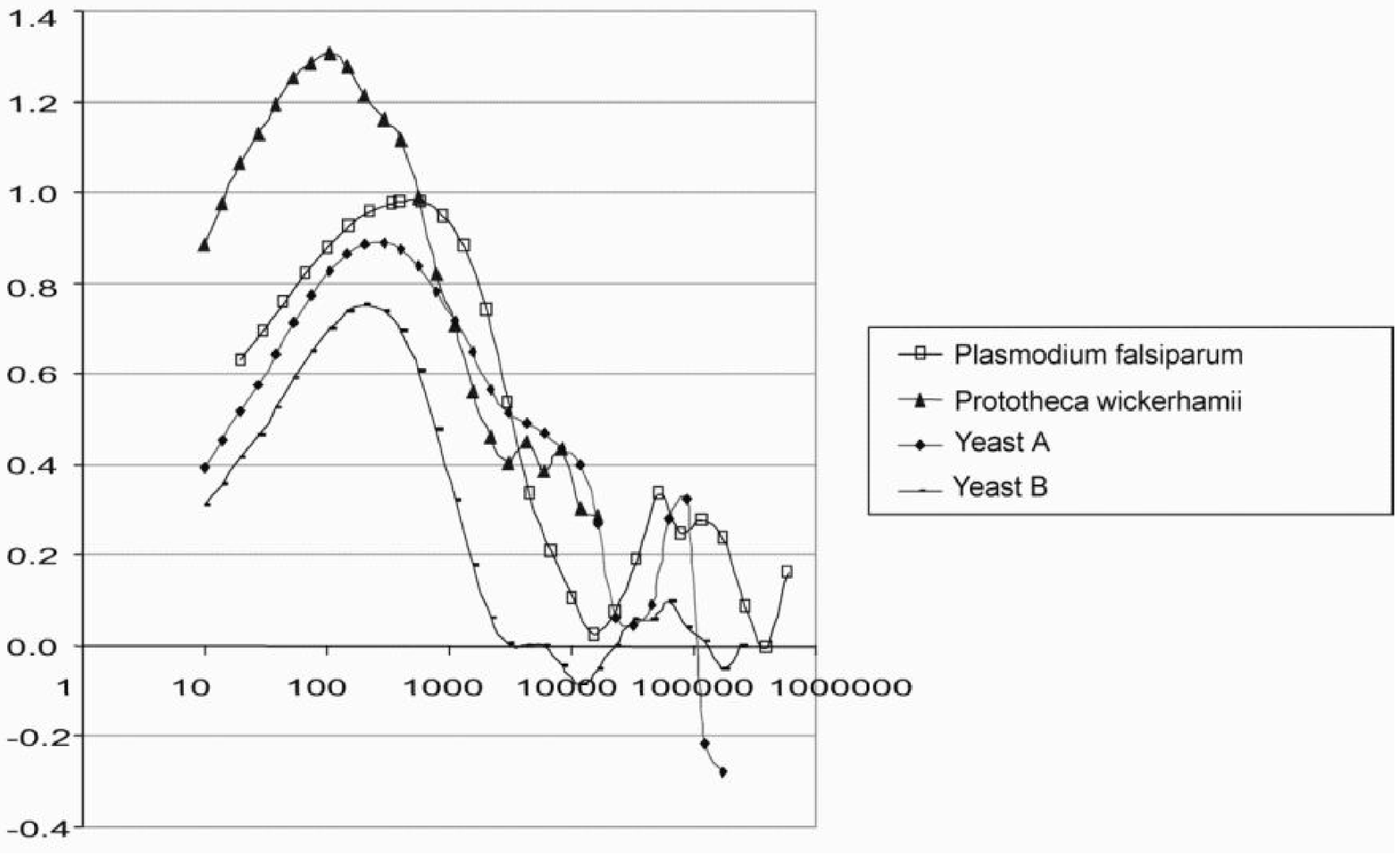}a)
\includegraphics[width=70mm, height=60mm]{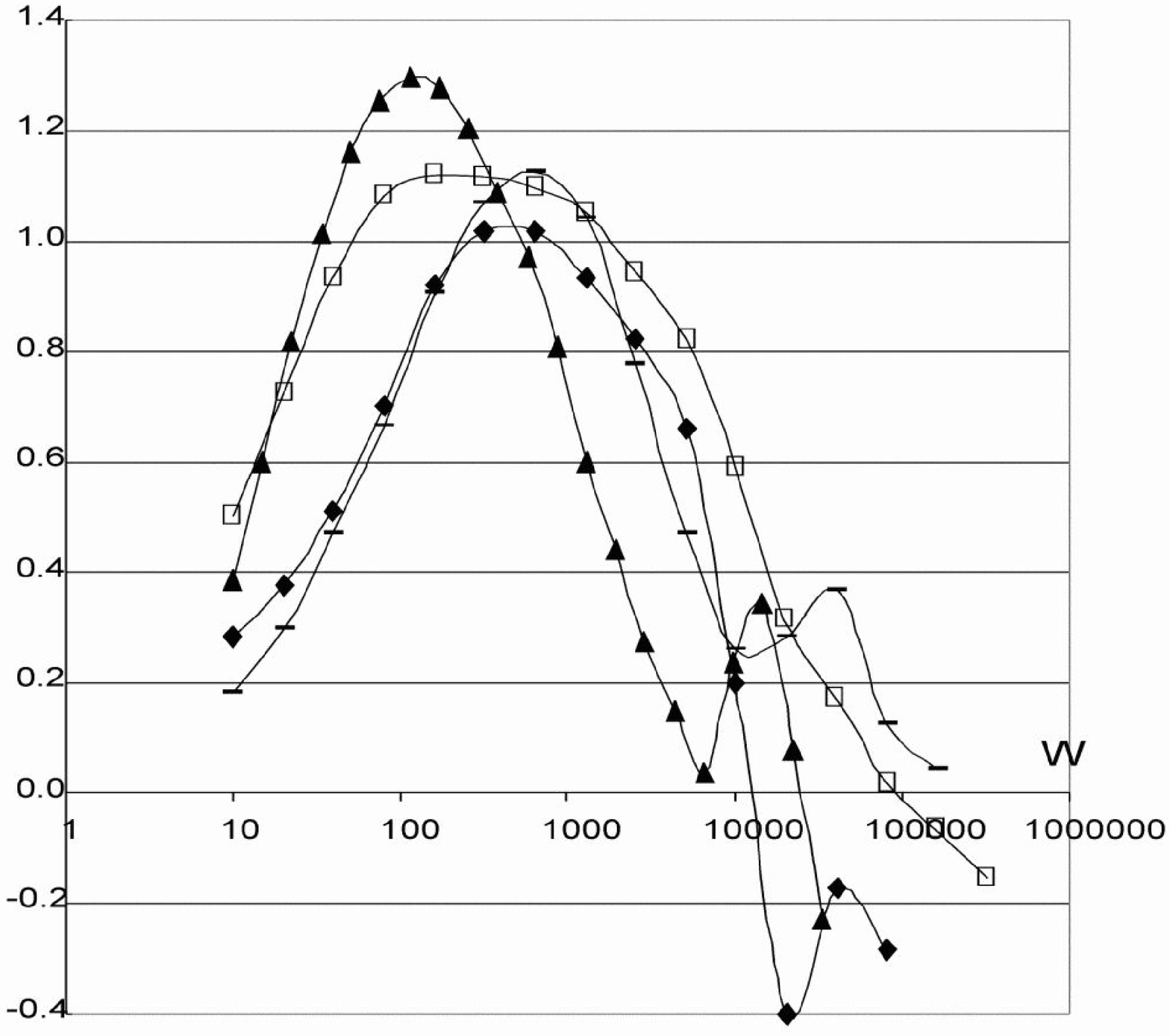}b)
\caption{Effectiveness of two measures (local GC-concentration
(a), mixing entropy (b)) for several genomes. Bimodal character of
graphs can be explained: first maximum is the difference of coding
and non-coding regions themselves, second is statistical
difference of long regions (isochores)}
\end{figure}

Another promising approach is to partition the DNA sequence into
homogenious in some sense subsequences and to find such a way of
partitioning that corresponds to the "coding -- non-coding"
partition. In the works of Bernaola-Galvan et.al.\cite{Bernaola}
method of entropic segmentation was formulated that uses
difference in codon compositions in coding and non-coding regions.
The hypothesis is that codon composition in coding regions is
different from junk because of well-known fact of biasing in codon
usage.

Methods of gene finding use a variety of numerical characteristics
reflecting statistical regularities in some subsequence (window)
of DNA. Inphase hexamers seem to be the most effective single
measure (see, for example, \cite{Fickett96}). Calculation of
inphase hexamers based on division of given subsequence into
non-overlapping triplets and counting for every triplet dicodon
occurences starting from the first, second and third position in
the triplet.

In this work we introduce a method for identification of
protein-coding regions in DNA that uses notion of distinguished
coding phase. We try to explain the reasons for measures which use
in a way the idea of coding phase (including such measures as
inphase hexamers, assymetry, entropy etc., see
\cite{Fickett96,GorbanSAGI01} for definitions) to be useful in the
methods of identification of protein coding regions. Using the
idea in the explicit form we formulate procedure for
identification of protein-coding regions in a self-organizing
manner.

Let's take arbitrary subsequence of DNA (below we think of both
DNA strands as of one chain, not touching problem of possible
genes overlapping) and divide it into non-overlapping triplets in
three different ways, starting from the first, second and third
basepair in the window. We get 3 distributions of triplet
frequencies $f_{ijk}^{(1)}$, $f_{ijk}^{(2)}$, $f_{ijk}^{(3)}$,
$i,j,k \in \{A,C,G,T\}$. Then we can consider mixed distribution
$f_{ijk}^{(s)}=\frac{1}{3}(f_{ijk}^{(1)}+f_{ijk}^{(2)}+f_{ijk}^{(3)}).$

If we suppose that given subsequence is protein coding and
homogeneous (without introns) then one of the three distributions
is the real codon distribution and it defines distinguished coding
phase. No matter the distribution is, other two are derivable from
it (provided fixed mixed distribution and under several
assumptions) and, generally speaking, should be different. This
happens due to the fact that the distribution of triplets in
coding phase is strictly conserved in the process of evolution.

The situation is fairly opposite in non-coding regions. Due to
allowable operations of deleting and inserting a basepair in
sequence, all three distributions are expected to be mixed and
equal to $f_{ijk}^{(s)}$.

It is worth noticing, that distributions $f_{ijk}^{(1)}$,
$f_{ijk}^{(2)}$, $f_{ijk}^{(3)}$ are projections of distribution
of pentamers $p_{ijklm}$, $i,j,k,l,m \in \{A,C,G,T\}$ which
counted from every third position starting from the first basepair
in a window. It means that information contained in distribution
of pentamers  seems to be sufficient for prediction of coding
regions with the same accuracy as using hexamers (but requires
shorter subsequence to evaluate frequencies).

Another interesting note is that GC-concentration in a window is
the linear function of the frequencies of triplet distribution in
any phase. It means that in the space of triplet (or pentamer,
hexamer etc.) frequencies gradient of this functional determines a
distinguished direction along what the separation of coding and
non-coding windows is good. Really, it is well-known fact that
coding regions are GC-rich comparing to non-coding. We will show
below that the difference in GC-concentration between coding and
non-coding regions is most contrast at the scales comparable to
the average gene length in genome.

To implement the simple idea of distinguished phase as a procedure
for identification of protein coding regions in DNA, first we
investigated dependence of effectiveness of two simple measures on
the length of sliding window. It was done on the known genome
anotations and it was shown that the dependence has bimodal
character and is not very strong in some range of window lengths.

\begin{figure}
\centering{
\includegraphics[width=50mm, height=40mm]{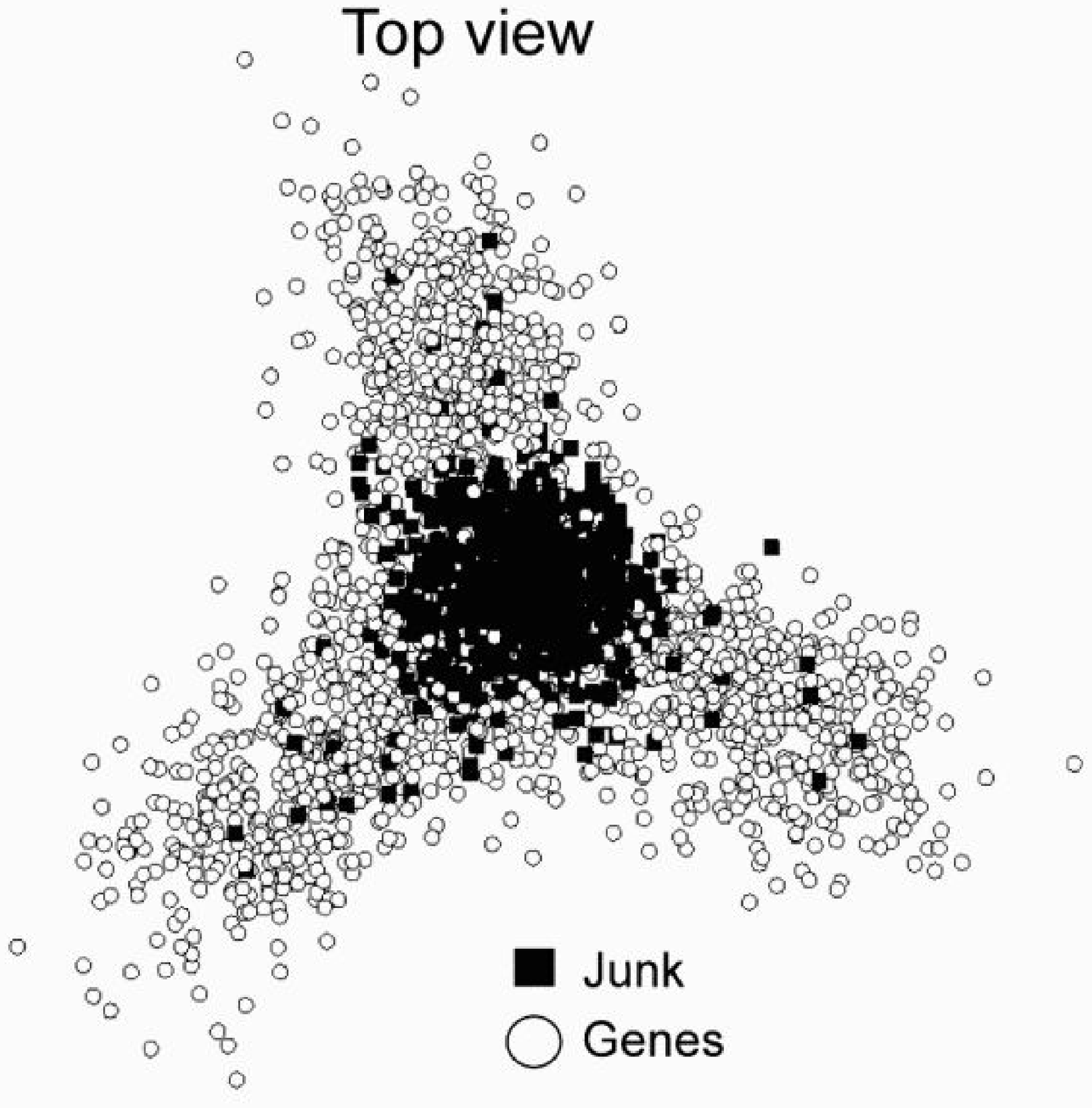}
\includegraphics[width=50mm, height=40mm]{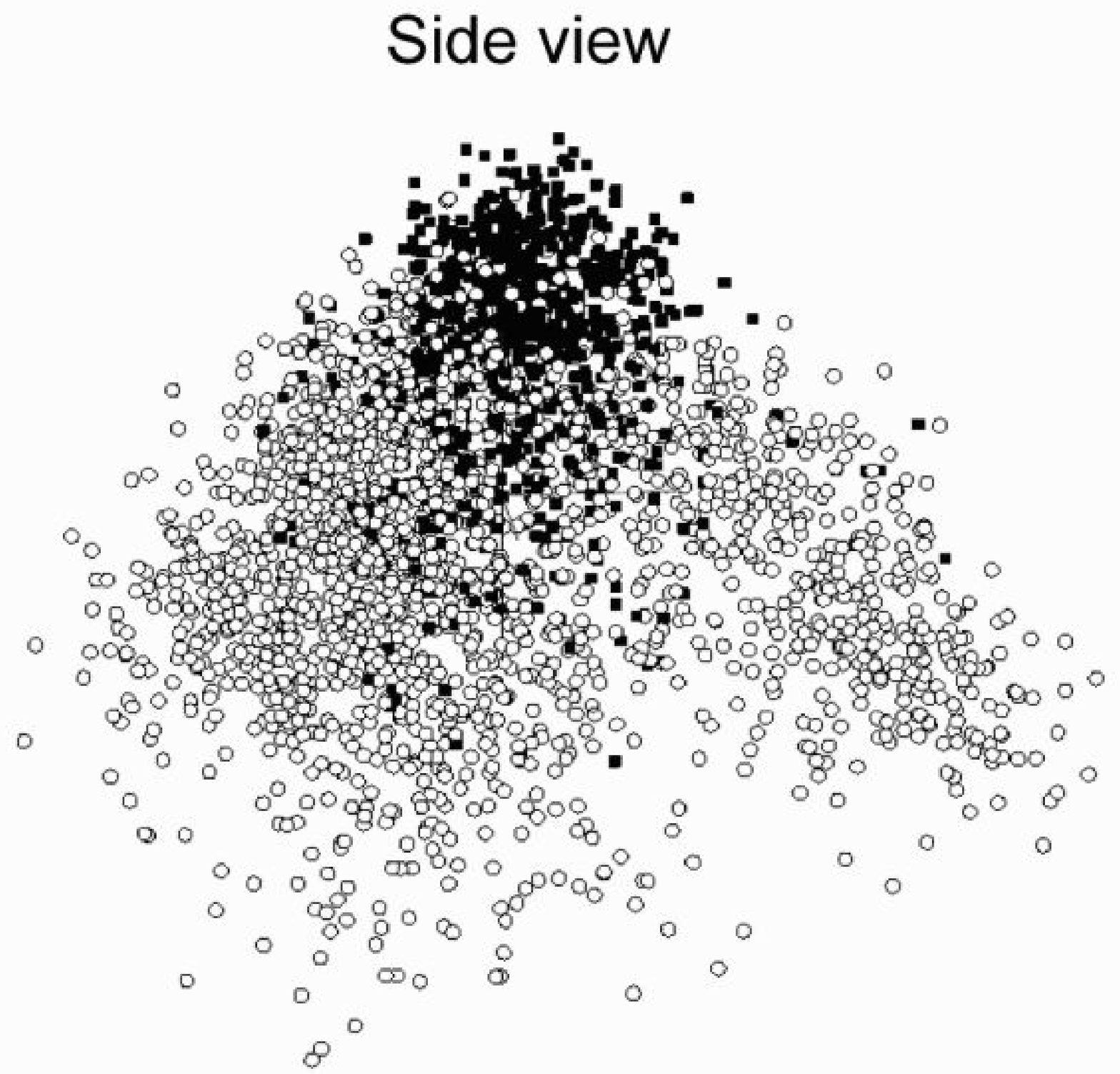}
\caption{Distribution of triplet frequencies in the space of first
three principal components for P.Wickerhamii}}
\end{figure}

\begin{figure}
\centering{
\includegraphics[width=50mm, height=5mm]{empt.ps}
\includegraphics[width=50mm, height=40mm]{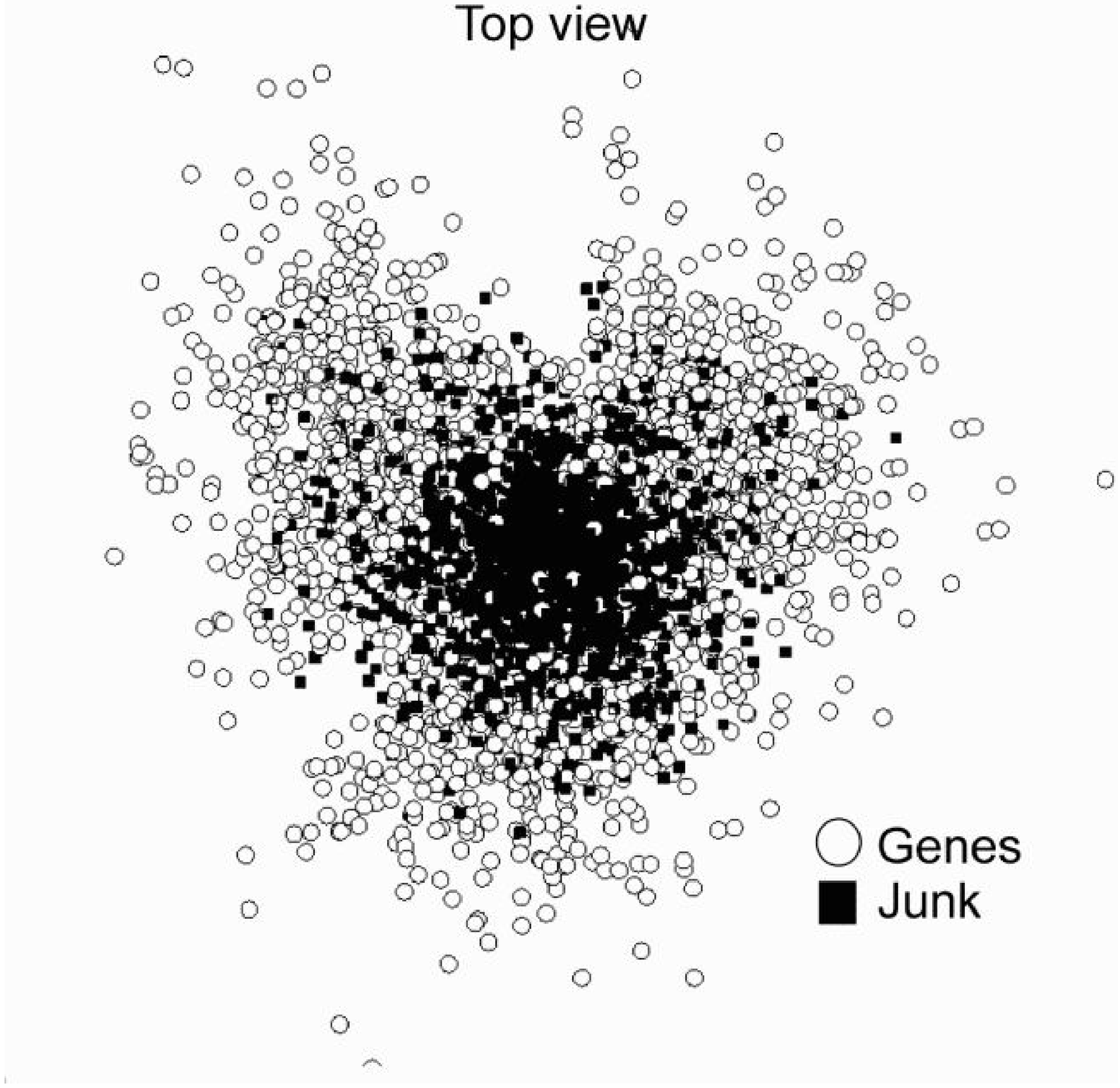}
\includegraphics[width=50mm, height=40mm]{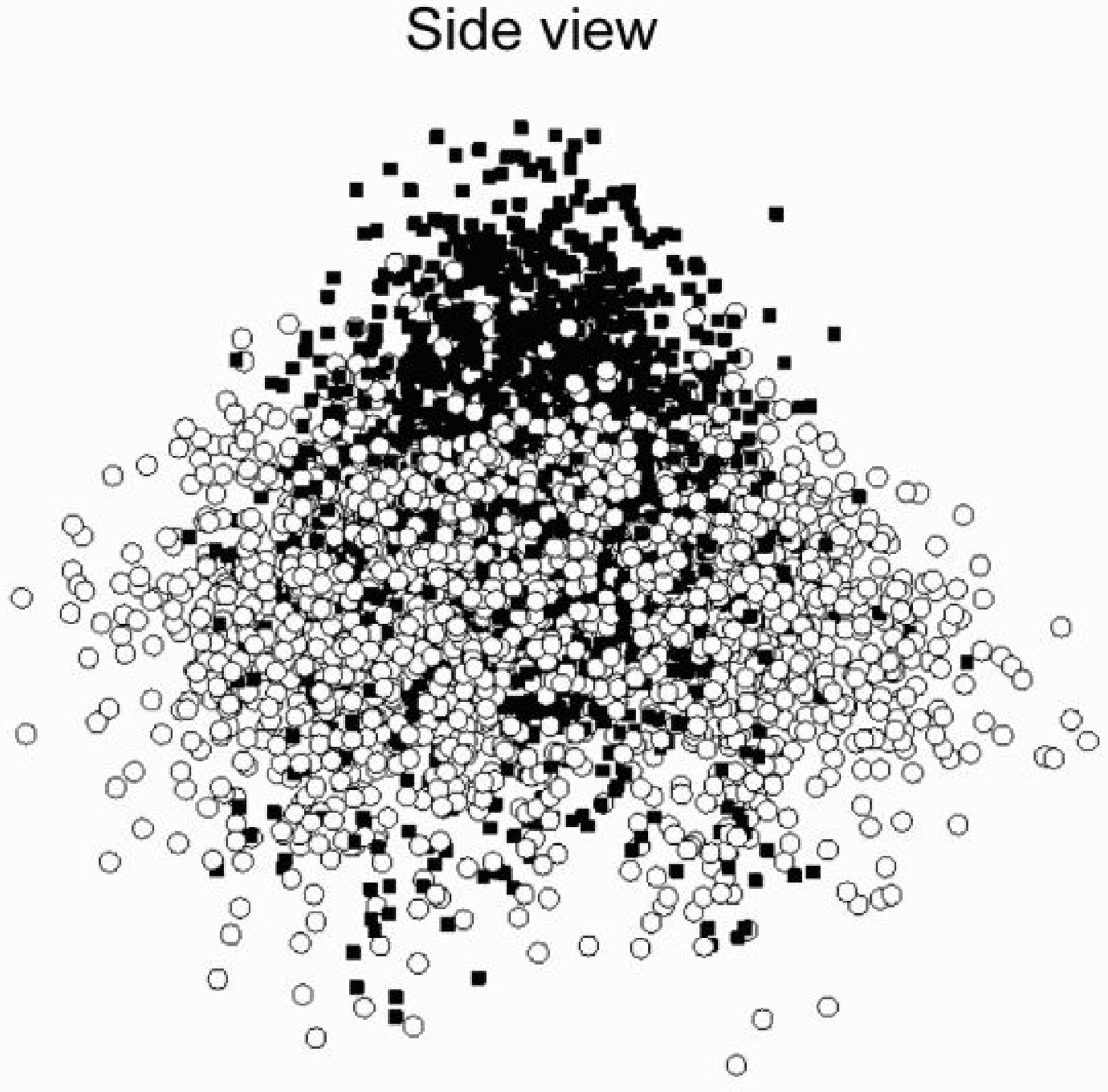}
\caption{Distribution of triplet frequencies in the space of first
three principal components for S.Cerevisiae III}}
\end{figure}

Suppose every nucleotide in one strand of DNA to be "coding" or
"non-coding". Then assume that this property depends on some
measure calculated over the whole window with length W, centered
in the position of the nucleotide. We may evaluate the
effectiveness of this measure for separation of coding subset
${\bf G}$ and non-coding subset ${\bf J}$ in the set of all
nucleotides. Let $A_W(i)$ be our measure calculated for the $i$-th
position and

$$\Delta_W=\frac{ \frac{1}{W} (\sum_{i \in {\bf G}} A_W(i)-\sum_{i
\in {\bf J}} A_W(i))}{\sqrt{DA_W(i)}}$$

be a measure of effectiveness, where $D$ is dispersion of $A_W(i)$
over the whole set $\{{\bf G},{\bf J}\}$. In figures 1(a), 1(b)
dependence of $\Delta_W$ on W for two measures and several genomes
is shown. First (fig.2(a)) is local concentration of GC-bonds in a
window. Second (fig.2(b)) is so called "mixing entropy"
$S_M=\frac{1}{3}(3S-S^{(1)}-S^{(2)}-S^{(3)})$, where
$S=-\sum_{ijk}f_{ijk}^{(s)}\ln f_{ijk}^{(s)}$,
$S^{(m)}=-\sum_{ijk}f_{ijk}^{(m)}\ln f_{ijk}^{(m)}$. It is clear
that sets $\bf J$ and $\bf G$ can be separated with confidence
(with enormous number of points we have difference of two mean
values more than one standard deviation) and effectiveness of
$S_M$ measure seems to be better then GC-average. An optimal
window length for calculating the measures is about 400 bp for
S.Cerevisiae and P.Falsiparum genomes and about 120 bp in the case
of short mitochondrial genome.

Using these values we constructed a finite set of points in
64-dimensional space of triplet frequencies, each point
corresponds to the frequencies distribution $f^{(1)}$ of
non-overlapping triplets with phase 1 (starts from the first
basepair in window). Then coordinates of points in the set
$X=\{x_i\}, i=1...N$ were normalized on unit standard deviation:
$$\tilde{x}_{ij}=\frac{x_{ij}-\bar{x}_j}{\sigma_j},$$ where
$x_{ij}$ is the $j$-th coordinate of the $i$-th point and
$\bar{x}_j$, $\sigma_j$ are mean value and standard deviation of
the $j$-th coordinate.

\begin{figure}
\centering{
\includegraphics[width=70mm, height=35mm]{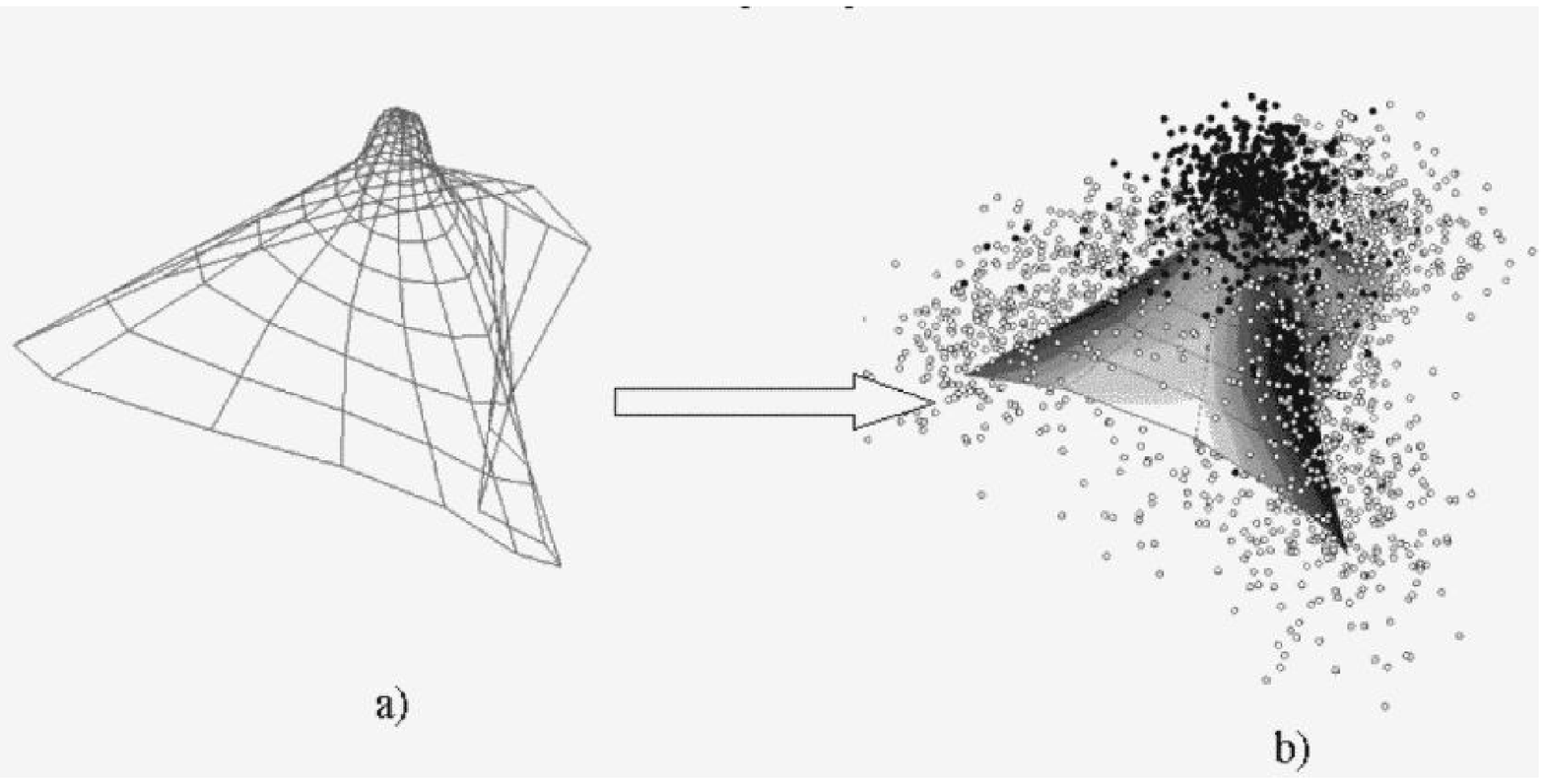}
\includegraphics[width=45mm, height=45mm]{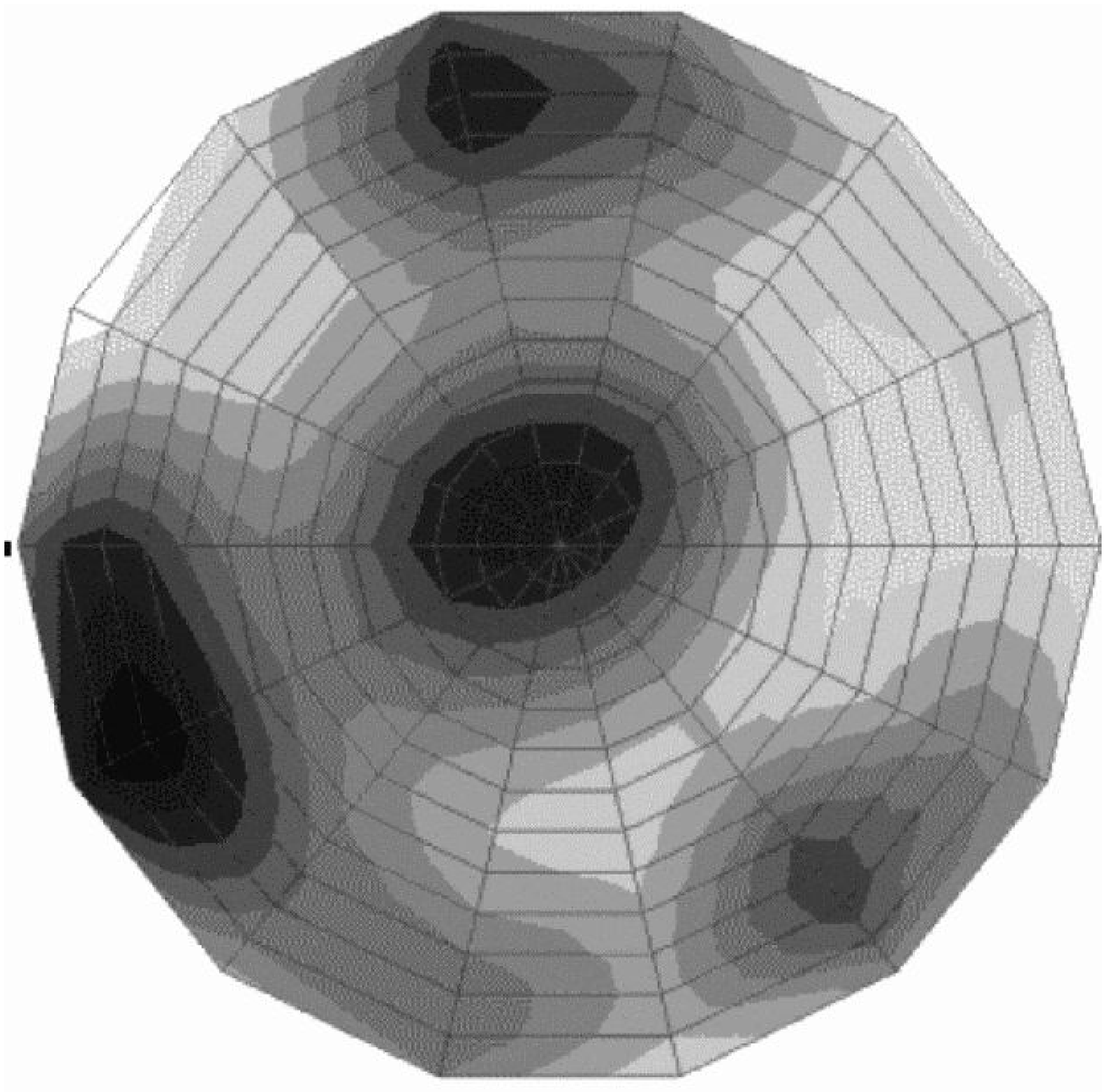}
c) \caption{Two-dimensional visualization of distribution density
using method of elastic map: a) form of constructed elastic map;
b) position of the map in data space (projection on the first
three principal components); c) resulting picture of estimation of
density distribution.}}
\end{figure}

The set of normalized vectors $\tilde{x}_i$ was projected into the
subspace spanned by the first three principal components of the
distribution and visualized with showing known separation for
coding and non-coding nucleotides (see fig.2,3). The distribution
has bullet-like structure with a kernel corresponding to the
non-coding regions (where there is no distinguished phase) and
three tails which correspond to the three possible shifts of real
codon distribution to the phase of test triplets in a window.

To visualize density of the distribution more advanced technology
was used named "method of elastic maps"(see
\cite{GorbanRossiev99,Zinovyev00,GorbanMash00,GorbanVisu01}). The
method of elastic maps just like self-organizing maps
\cite{Kohonen97} constructs point approximation to the non-linear
principal 2D-surface using minimization of elastic energy
functional that consist of three parts describing "node -- data
points" attraction and energies of stretching and binding of the
net of nodes with appropriate topology. More isometric than in SOM
net of nodes allows to construct piece-wise linear 2D-manifold and
to project data points in a piece-wise linear fashion onto it,
then using the manifold as a 2D screen for visualization purposes.
In our case we initialized the net on the 2D-hemisphere put into
multidimensional space. After that it was deformed using algorithm
of construction elastic net for the optimal approximation and the
coloring was used to visualize the resulting density of
projections of data points (more precisely, its non-parametric
estimation). The distribution of data points has four clusters
(fig.4), corresponding to the non-coding regions (central cluster)
and protein coding (three peripherial clusters).

Using this fact the procedure for unsupervised prediction of
protein coding regions may be formulated. We construct
distribution of triplet frequencies just as we did it (using some
suboptimal value of window length) and then cluster it for 4
clusters, using appropriate clustering algorithm. It gives
separation of all nucleotides into non-coding (0-phase) and
protein-coding (1,2,3-phase).

\begin{figure}
\centering{
\includegraphics[width=85mm, height=50mm]{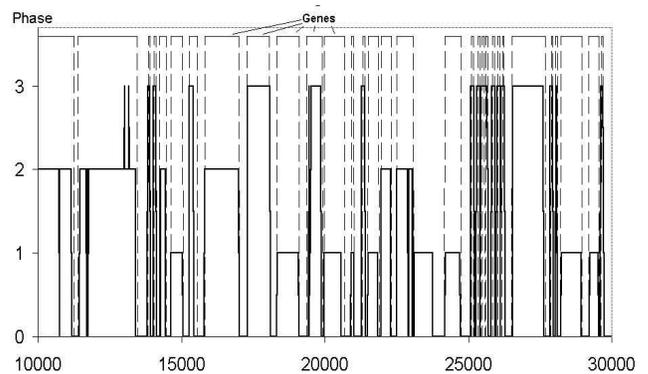}
\caption{Prediction of protein-coding regions using clustering in
the space of triplet frequencies for P.Wickerhamii. X-axis is
basepair position in sequence, Y-axis is number of cluster (coding
phase).}}
\end{figure}

\begin{figure}
\centering{
\includegraphics[width=85mm, height=50mm]{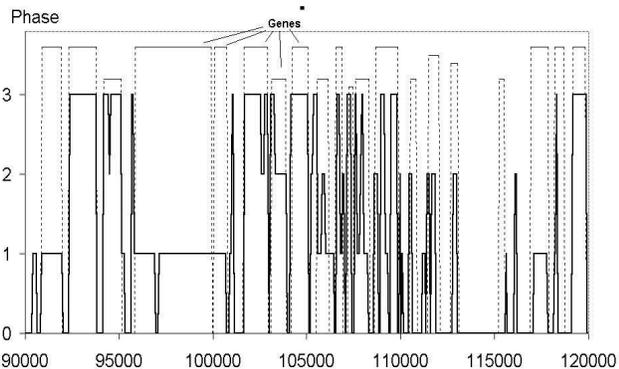}
\caption{Prediction of protein-coding regions using clustering in
the space of triplet frequencies for S.Cerevisiae III. X-axis is
basepair position in sequence, Y-axis is number of cluster (coding
phase). Dotted line shows positions of ORFs, the height of bar
corresponds to the confidence of gene presence (highest bars are
experimentally discovered genes.) }}
\end{figure}

We used simplest method of K-means for clustering and found that
separation of nucleotides in investigated genomes relates to the
known data with accuracy from 65\% up to 85\% (calculating
accuracy as percentage of correctly predicted nucleotides - coding
and non-coding). Though these results are comparable with
performance of gene-finders used in real practice \cite{Claverie},
more advanced techniques for clustering promise better results.
Fragments of the resulting graphs of phase (that is actually
cluster number) of sliding window (calculated through every 3 bp)
are shown in fig.5,6.

So, in this paper it was demonstrated that simple notion of
distinguished coding phase in three possible distributions of
triplets in a window of DNA lays as background in various methods
of gene finding. Visualization of the set of sliding windows in
the space of triplet frequencies shows symmetric bullet-like
structure. Linear dimensions of the structure are determined by
amplitudes of two measures: local GC-concentration and mixing
entropy.

These two measures have maximum of their effectiveness for
separating coding and non-coding regions in the same quite wide
range of window lengths (relating to the average length of gene).
As average mixing entropy measure is more effective, but it can
separate only protein-coding regions, while effectiveness of
GC-concentration does not depend on the type of the coding region.

Analysis shows that distribution of windows of DNA in triplet
frequencies space forms 4 clusters (central one for junk region,
where there is no coding phase, and 3 side ones for three possible
phase shifts). Though this clustering is not very compact, it may
be used for gene-finding without any learning dataset.

\begin{acknowledgments}
Our efforts were inspired by attention of Prof. M.Gromov to the
work. We are thankful to Prof. A.Carbone (IHES) for stimulating
discussion and help. The paper of A.Carbone and M.Gromov
\cite{Gromov01} was a good inspiring example for us of how
molecular biology can be thought from a mathematical angle.
\end{acknowledgments}

\end{document}